\documentclass[a4paper,11pt]{article}
\pdfoutput=1 

\usepackage{jcappub} 

\usepackage[T1]{fontenc} 

\usepackage{aas_macros}
\usepackage{booktabs}
\usepackage[separate-uncertainty]{siunitx}
\usepackage{subcaption}

\newcommand{\dlum}{\ensuremath{d_\text{L}}}

\newcommand{\dang}{\ensuremath{d_\text{A}}}
\newcommand{\dlrec}{\ensuremath{d_\text{L}^\mathrm{rec}}}
\newcommand{\lcdm}{$\Lambda$CDM}

\newcommand{\Ok}{\ensuremath{\mathcal{O}_k}}

\newcommand{\Omm}{\ensuremath{\Omega_\text{m}}}
\newcommand{\Omb}{\ensuremath{\Omega_\text{b}}}
\newcommand{\Oml}{\ensuremath{\Omega_\Lambda}}
\newcommand{\Omk}{\ensuremath{\Omega_k}}
\newcommand{\Omr}{\ensuremath{\Omega_\text{r}}}

\newcommand{\tens}[1] {\ensuremath{\mathbf{#1}}} %
\newcommand{\rd}{\ensuremath{r_\text{d}}} 
\newcommand{\diff} {\ensuremath{\mathrm{d}}}

\newcommand{\DD}{\ensuremath{\mathcal{D}}}

\newcommand{\chsq}{\ensuremath{\chi^2}}

\author[a]{Benjamin L'Huillier}
\author[a,b]{and Arman Shafieloo}

\affiliation[a]{Korea   Astronomy   and   Space   Science   Institute,
  Yuseong-gu, 776 Daedeok daero, Daejeon, Korea}
\affiliation[b]{University of  Science and Technology,  Yuseong-gu 217
  Gajeong-ro, Daejeon, Korea} 

\emailAdd{benjamin@kasi.re.kr}
\emailAdd{shafieloo@kasi.re.kr}

\abstract{%
  Using measurements of $H(z)$ and $\dang(z)$ from the Baryon
  Oscillation   Spectroscopic  Survey   (BOSS)  DR12   and  luminosity
  distances from  the Joint  Lightcurve Analysis (JLA)  compilation of
  supernovae (SN), we measure $H_0\rd$ without any model assumption.
    Our measurement  of
    $H_0  \rd  =  (\num{10033.20 }  ^{+333.10}_{-371.81}  \,(\text{SN})
    \num{\pm 128.19} \,(\text{BAO}))\, \si{km.s^{-1}}$ 
    is  consistent with  Planck  constrains for  the  flat \lcdm\  model. 
    We also report that higher  expansion history rates $h(z)$ (among the
    possibilities)  as well  as lower-bound  values of  $H_0\rd$ result  in
    better internal consistency among the independent  data ($H(z)\rd$ and
    $\dang(z)/\rd$ from  BAO at  $z=0.32$ and  $z=0.57$ and  \dlum\ from
    JLA) we used in this work.
    This can be interpreted as an interesting and independent support of
    Planck cosmology without using any cosmic microwave background data.  
    We    then     combine    these    observables    to     test    the
    Friedmann--Lema\^itre--Robertson--Walker   (FLRW)  metric   and  the
    flatness  of  the  Universe  in   a  model-independent  way  at  two
    redshifts, namely 0.32 and 0.57, by introducing a new diagnostic
    for flat-FLRW,  $\Theta(z)$, which  only depends on  observables of
    BAO and SN data. 
    Our   results  are   consistent  with   a  flat-FLRW
    Universe within $2\sigma$.
  } 

\title{Model-independent test of the FLRW  metric, the flatness of the
  Universe, and non-local estimation of $H_0\rd$}


\arxivnumber{1606.06832}

\begin{document}

\maketitle 

\flushbottom

\section{Introduction}

The current concordance model of  the Universe relies on the important
assumptions that  the Universe is  isotropic and homogeneous  on large
scale, and that gravity is described by General Relativity.
Under these assumptions, the solution to Einstein's field equations is
the Friedmann--Lema\^itre--Robertson--Walker (FLRW) metric.
Considering other aspects of the  concordance model, such as assumption
of  the  power-law  form  of  the  primordial  spectrum  and  assuming
cosmological constant  as dark energy,  we can make predictions  on the
behaviour of the  Universe in different contexts and  confront it with
cosmological observations to constrain its six basic parameters. 
However,  testing  different  assumptions  of the  standard  model  of
cosmology, including its curvature and  metric, is still of primordial
importance \cite{2008PhRvL.101a1301C, 2009PhRvD..80l3512W,
2010PhRvD..81h3537S,
2011arXiv1102.4485M, 2012ApJ...748..111W,2014ApJ...789L..15L,
2014PhRvD..90b3012S,
2015PhRvL.115j1301R, 2015PhRvD..91f3506Z, 2016IJMPD..2530007B}.

Supernovae (SN) can be used to trace the expansion history of the Universe,
and have  revealed its recent  acceleration \cite{1998Natur.391...51P,
  1998AJ....116.1009R, 2007ApJ...659...98R}.  
In addition, the  baryonic acoustic oscillations (BAO) can  probe the growth
of structures \cite{2005ApJ...633..560E, 2016MNRAS.457.1770C,
   2016MNRAS.tmp..930G} by measuring $H(z)\rd$ and $\dang(z)/\rd$, where 
 \begin{equation}
  \rd = \frac c {\sqrt{3}} \int_0^{1/(1+z_\text{drag})} \frac{\diff a}
      {a^2 H(a) \sqrt{1+\frac {3\Omb}{4\Omr}}},
\end{equation}
is the sound horizon at the  drag epoch, and \Omb\ and \Omr\ are
  the baryon and radiation density parameters at $z=0$.

In this work we use a non-parametric approach to derive  $H_0\rd$,
combining  independent data from  supernovae  and   baryon  acoustic
oscillations,  and test  the consistency  of the  results. Using  some
local measurements of $H_0$ we can then estimate the value of $\rd$ in
a non-parametric and model-independent manner.
Measuring \rd\ model-independently avoids to bias the results toward a
particular     cosmological      model     \citep{2012MNRAS.426.1280S,
  2014PhRvL.113x1302H}.
We should also note that in our paper we use a non-parametric approach
to reconstruct the expansion history of the universe and this makes our
results even free from assuming any particular parametric form.
We then  test the  FLRW metric  and flatness of  the Universe  using a
non-parametric  and   model-independent  approach  depending   on  the
observables directly.

Section \ref{sec:method} describes the method  used in this study, and
the results are presented in \S~\ref{sec:res}.
In \S~\ref{sec:ccl} we discuss about our results and conclude.

\section{Method}

\label{sec:method}

In  a FLRW  metric, the  curvature parameter  $\Omk$ is  constant with
redshift (its  associated energy varies  like $\Omk (1+z)^2$)  and the
luminosity distance can be written as \cite{1972gcpa.book.....W}:
\begin{align}
	\label{eq:dl}
  d_\text {L}(z)  & = (1+z)^2 d_\text{A}(z)  = \frac {(1+z)d_\text{H}}
  {\sqrt{  -\Omk }}  \sin  \left( \sqrt  {-\Omk} \int_0^z  \frac{\diff
    z'}{h(z')}\right), \intertext{where} 
  d_\text{H} &= \frac c {H_0}
  \intertext{is  the  Hubble  distance, $d_\text{A}$  is  the  angular
    diameter distance, and}
  h^2(z)    &=    {\Omm(1+z)^3   +    \Omk(1+z)^2    +
    \Omega_{\mathrm{DE}}\, \mathrm{exp}{\left(3\int_0^z \frac{1+w(z')}
                                {1+z'}  \diff z'\right)}}
  \intertext{for a FLRW universe where the equation
            of state of  dark energy is $w  (z)=P/\rho$ ($w(z)=-1$ for
            the cosmological             constant). 
            It is also useful to  work with the dimensionless comoving
            distance defined as} 
  \DD(z) &= \frac 1 {(1+z) d_\text{H}} d_\mathrm{L}(z) = \frac 1
     {\sqrt{-\Omk}} \sin  \left( \sqrt  {-\Omk} \int_0^z  \frac{ \diff
       z'} {h(z')}\right)\label{eq:dd}.
\end{align}



The main idea  here -- to test the curvature, FLRW  metric, estimate $H_0\rd$,  and  test the  internal  consistency  between
different  observations --  is  to   obtain  $\DD(z)$  -  $\DD'(z)$  and
$d_\text{A}(z)/\rd$  -   $H(z)\rd$  from  independent   measurements  without
assuming a cosmological model.
In order to  get $\DD'(z)$, we can evaluate $\DD(z)$  at all redshifts
using the distance modulus of supernovae and calculate its derivative. 
In this work, we use the JLA sample \citep{2014A&A...568A..22B}.  
In
\citep{2006MNRAS.366.1081S, 2007MNRAS.380.1573S, 2010PhRvD..81h3537S},
the  authors introduced  a  non-parametric method  to reconstruct  the
luminosity distances from supernovae data by iteratively smoothing the
residuals.
The log-normal smoothing kernel and iterative approach used in
\citep{2006MNRAS.366.1081S,  2007MNRAS.380.1573S, 2010PhRvD..81h3537S}
has shown the effectiveness of  the method in direct reconstruction as
it has already been implemented succesfully in different contexts.
Non-parametric   methods  can   be  very   useful  because   they  are
model-independent, which  enables them to look  for unexpected features
in the data beyond the flexibility of parametric approaches.
The actual quantity recovered by this method is 
\cite{2010PhRvD..81h3537S}
\begin{equation}
\label{eq:drec}
\dlrec(z) = H_0/c \, \dlum(z) = (1+z)\,\DD(z). 
\end{equation}

On the other  hand, $d_\text{A}(z)/\rd$ and $H(z)\rd$  can be measured
directly  by the  baryonic acoustic  oscillations,  for instance
from the Baryon Oscillation Spectroscopic Survey (BOSS) DR12
\citep{2016MNRAS.457.1770C}. 
We combine all these results for the purposes of this work. 
 
\subsection{Smoothing}
\label{sec:smoothing}
In our procedure we go through the following steps:
\begin{enumerate}
\item
  Start with different initial guess models, 
\item
  Iteratively smooth the  data with a kernel similar to  what has been
  proposed before, using $N_\text{iter}$ iterations, 
\item
  Only keep those  reconstructions that yield a better  $\chsq$ than a
  reference model.  
\end{enumerate}

We use the  best-fit flat \lcdm\ as our reference  model. Note that we
do not consider these reconstructions  to be necessarily more probable
than  the reference  model.  We  use the  reference  model  only as  a
criterion to  accept a  non-exhaustive sample of  reconstructions with
plausible and probable expansion histories.  

In order to  explore the allowed \DD, we start  with different initial
guesses:  best-fit  flat-\lcdm\,  open  \lcdm,  $w$CDM,  standard  CDM
($\Omm=1$), open  CDM ($\Omm+\Omk=1$),  empty universe  ($\Omk=1$), de
Sitter  universe ($\Oml=1$),  as  well as  flat-\lcdm\ universes  with
$\Omm=0.1,\dots,0.9$. 

We     followed    \cite{2006MNRAS.366.1081S,     2007MNRAS.380.1573S,
  2010PhRvD..81h3537S}  and used  a  log-normal kernel  to smooth  the
supernovae data, taking into account the errors in $\mu$.
We  start from  an initial  guess $\hat\mu_0(z_i)$  at the  input data
$z_i$ (see previous section).
The smooth distance modulus at iteration $n+1$ is then calculated by
\begin{align}
  \label{eq:smooth}
  \hat\mu_{n+1}(z)       &=       \hat\mu_n(z)      +       N(z)
  \sum_i{  \left(  \frac{  \mu(z_i)   -  \hat  \mu_n(z_i)}  {\sigma_i^2}
    \exp{ \left( - \frac{ \ln^2 \left(  \frac{ 1+z_i } {1+z} \right) }
      {2 \Delta^2} \right)} \right)}\\ 
  N^{-1}(z)  &=  \sum_i{ \left(  \frac  1  {\sigma_i^2} \exp{  \left(  -
               \frac{  \ln^2  \left(  \frac{1+z_i} {1+z}  \right)}  {2
                 \Delta^2} \right)} \right)},
\end{align}
where $\hat\mu_{n}(z)$  is the  reconstructed distance modulus  at any
redshift  $z$,  while  $\mu(z_i)$  and  $\sigma_i$  are  the  measured
distance modulus 
and  its  associated  error  at   redshift  $z_i$;  and  $N(z)$  is  a
normalisation factor.  
At each iteration, we calculate the \chsq\ defined as
\begin{align}
  \chsq_n & = (\mu - {\hat\mu_n})^{\mathrm{T}}
  \tens{C}^{-1} (\mu - {\hat\mu_n}), 
\end{align}
where $\tens C$ is the data covariance matrix.

Note  that   the  supernovae   data  provide   us  with   $\mu(z)  =
5\log_{10}\dlum(z)+25 = m(z)-\mathcal{M}$, which can be rewritten as $\mu(z) =
5\log_{10}(c\dlum(z)/H_0) + \mu_0$, where $\dlum(z)$ is 
expressed in Mpc, and $\mathcal{M}$ is the absolute magnitude.

Following \cite{2010PhRvD..81h3537S},  we used $\Delta=0.3$  since the
number of supernovae and their uncertainties  are of the same order as
in \cite{2010PhRvD..81h3537S}. 
In practice, we  used $N_\text{iter} = 200$, and  keep only iterations
that yield a better \chsq\ than the reference model.
We used the best-fit flat \lcdm\ as our reference model. This  leaves us  
with  about 1600  trajectories of  \DD\  with better  \chsq\ than  the
reference model. 

In order to obtain 
$\DD'(z)=\diff \DD(z)/\diff  z = \diff (\dlrec(z)/(1+z))/\diff  z$, we
numerically calculate the derivative of $\DD$ and also  impose $\DD'(z=0)=1$.
 
\subsection{Error propagation}

To combine the results from smoothing  with the BAO data, we propagate
the errors  to combine  the uncertainties  from reconstruction  of the
expansion history  from supernovae data  and the uncertainties  of the
$H(z)$ from BAO data.
The errors on BAO ($H(z)\rd$ and $\dang/\rd$)   are  propagated
in the same way as in \cite{2010PhRvD..81h3537S}, taking into
account the  correlation between  them, and adding  external sources
(such as $H_0$) when needed. 
The errors  due to the  reconstructions from  the SNIa data  are taken
into account in the following way.
We  the median  as the  central  value, and  we used  the minimum  and
maximum derived values at each redshift to set the errors. 
Therefore we quote  two kinds of errors of  different nature, namely
SN and BAO.
We should note here again that  all reconstructions used here have a
better likelihood than the best-fit flat-\lcdm\ model (our reference
model). In other  words, the results derived in  this work represent
possibilities that  all have reasonable likelihood  (and better than
the best-fit $\Lambda$CDM model) given the data.

\section{Results}

\label{sec:res}

\subsection{Smooth reconstruction}

\begin{figure}
  \centering
  \includegraphics[width=\textwidth]{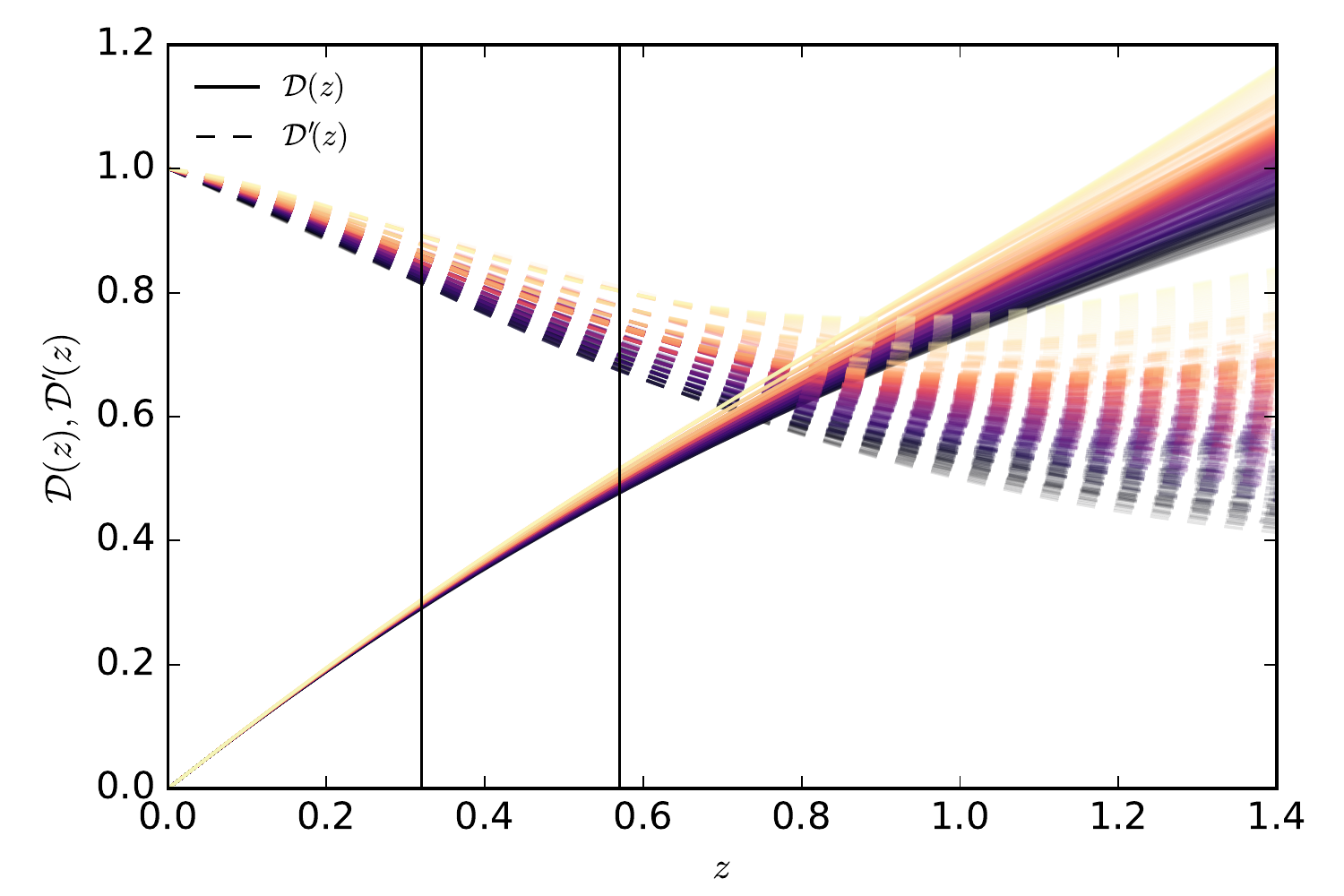}
  \caption{\label{fig:Dall}Reconstructed  $\DD(z)$  (solid lines)  and
    $\DD'(z)$ 
    (dashed lines) from JLA supernovae compilation. 
    All  lines shown  here have  a better  \chsq\ than  the reference
    model (best-fit  flat \lcdm).  Colors are  used to  show different
    reconstructed  $\DD(z)$   and  their   corresponding  (correlated)
    $\DD'(z)$.  
  }
\end{figure}

Figure  \ref{fig:Dall} shows  the  reconstructed \DD\  (solid lines)
and \DD' (dashed lines)  for several initial guesses.
All the curves shown here have a better \chsq\
than the reference model (best-fit \lcdm).
For  the  sake   of  visualization,  we  only   plotted  every  fourth
reconstruction. 
We colour-coded  the curves (according  to their value of  $\DD(z)$ at
$z=1.4$) since  each reconstructed $\DD(z)$ has  its own corresponding
(correlated) $\DD'(z)$.  
The agreement  in \DD\  at low-$z$  is very good,  owing to  the large
number of data  points, while at higher redshifts  the constraints are
less strong and different models start to deviate from each other. 
In our range of interest, $z  = 0.32$--$0.57$, the agreement is of the
order of 1\%.
However, \DD' shows some stronger deviation for different models which
is expected considering it to be a derivative of \DD.

\subsection{Estimation  of   $H_0\rd$  and  testing
  observational consistencies} 

\label{sec:H0rd}

In the  previous section, we  reconstructed $\DD'(z)$ at  any redshift
from the supernovae data.
Under assumption of flatness (eq.~\ref{eq:dd}), we have
$h(z) = 1/\DD'(z)$. 
We can combine it with $H(z)\rd$  obtained from the radial mode of the
BAO to estimate
\begin{equation}
  \label{eq:H0rdA}
  H_0\rd = \frac{H(z)\rd}{h(z)} = H(z)\rd\DD'(z).
\end{equation}
We will refer to this as method A.

Alternatively, combining eqs.~\eqref{eq:dl}  and \eqref{eq:dd}, we can
write 
\begin{align}
  \label{eq:H0rdB}
   H_0\rd &=  \frac{c \DD(z)\rd}{(1+z) \dang(z)}, 
\end{align}
where $\dang(z)/\rd$ comes from the transverse BAO \cite{2016MNRAS.457.1770C}.
We can calculate  $H_0\rd$ from the data at the  two redshifts of LOWZ
and CMASS.  
The main  interest of this  measurement is  the very high  accuracy of
$\dang(z)$.
Moreover, this  estimation of   $H_0\rd$  does not
assume flatness, since $\DD(z)$ is directly obtained from the smoothing.
This method will be referred to as method B.

We  can  thus  derive  $H_0\rd$  but  also,  we  can  test  the  internal
consistency of our  measurements where the derived $H_0\rd$  from method A
and B at both redshifts (0.32 and 0.57) should result in consistent
values.  

\begin{figure}
  \centering
  \includegraphics{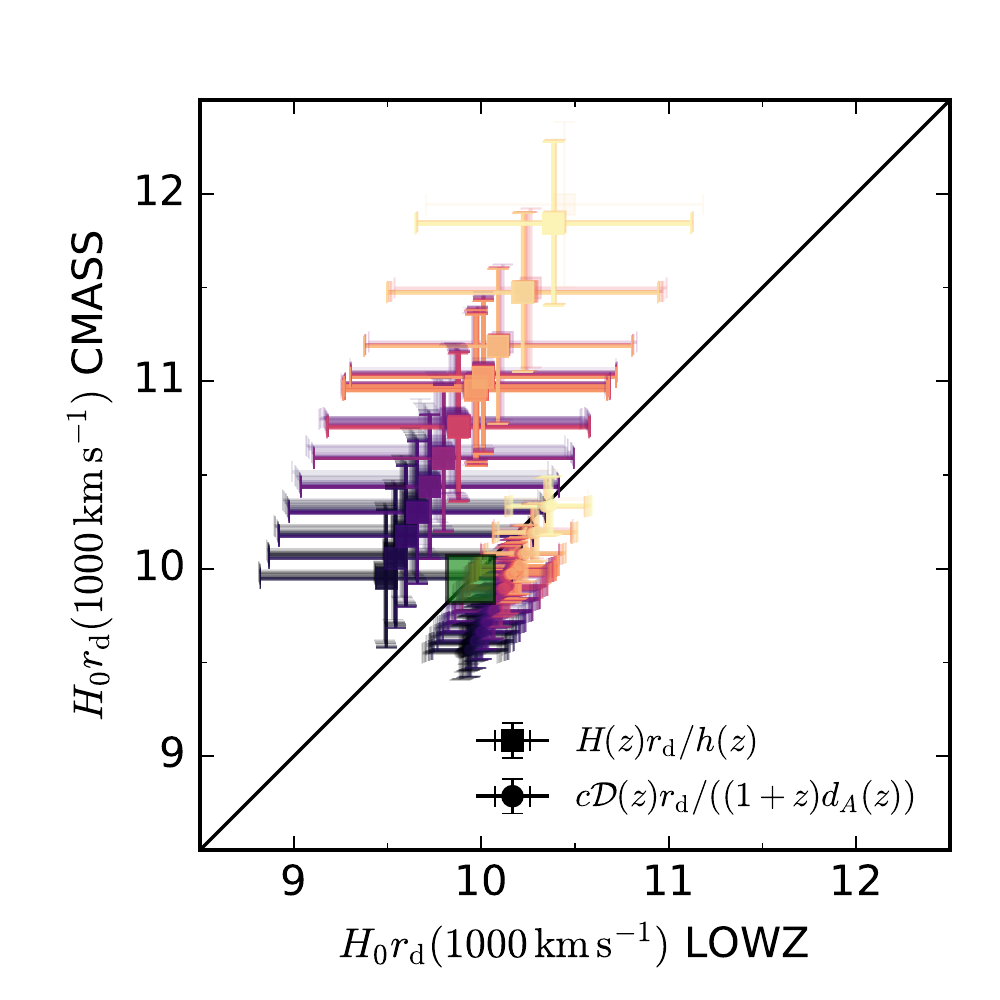}
  \caption{\label{fig:H0zboss}%
    $H_0\rd$ from LOWZ  versus CMASS from methods
    A (squares) and B (dots).
    The green square at $(\num{e4},\num{e4})$ \si{km.s^{-1}}
    shows the  $\pm 1\sigma$ region  centred around the  Planck 2015
    best-fit flat-\lcdm\ value for $H_0\rd$.
    Colour  code is  the  same as  Fig.~\ref{fig:Dall} for  different
    reconstructions of the expansion history from supernovae data.  
    One should  note that  there is a  better consistency  between all
    measurements for lower-bound values of $H_0\rd$
    and   higher   rate   expansion   histories   (darker   lines   in
    Fig.~\ref{fig:Dall}).  
  }
\end{figure}

Figure~\ref{fig:H0zboss} shows the value of
$H_0\rd$  from  LOWZ  ($x$-axis) versus  that  from
CMASS ($y$-axis) for method A (squares) and B (circles).
Each  point   is  given  by   one  reconstruction  of   $\DD(z)$  with
$\chsq<\chsq_\text{ref}$, while the error-bars come from the BAO.
The  colour-code  is  the  same  as  that  of  the  expansion  history
reconstruction shown in Fig.~\ref{fig:Dall}. 
One should  note that  results shown  with the  same colours  should be
consistent with each other.  
The     black,    diagonal     line    shows     the    loci     where
$H_0\rd$  from CMASS  and LOWZ  are equal,  and the
green square shows the $\pm  1\sigma$ region centred around the Planck
best-fit value.  

Method A (squares)  has larger error bars, due to the  larger errors on
$H(z)$ and also larger dispersion of $D'(z)$.
The     reconstructions     yielding      a     lower     value     of
$H_0\rd$ (black colour) 
are consistent  between CMASS and  LOWZ, while values  yielding larger
$H_0\rd$ (yellow) show inconsistency.  
Method B has smaller error bars, and larger values of
$H_0\rd\simeq \SI{10200}{km.s^{-1}}$  are consistent between  LOWZ and
CMASS, while lower values are slightly in tension.  

However, when considering both methods simultaneously, reconstructions
yielding higher  values of  $H_0\rd$ (yellow points)  appear to  be in
tension between the two methods. 
On the contrary, lower values of $H_0\rd\simeq\SI{9800}{km.s^{-1}}$
are preferred combining results of both methods at both redshifts.

We  also  calculated  the  best-fit  value  of  $H_0\rd$  using  the
flat-\lcdm\ Markov chains (TT, TE, EE, LowP, and lensing) from
\cite{2016A&A...594A..13P}: 
\begin{align}
  \label{eq:Planck15}
  H_0 \rd &= \SI{9944 +- 127.4}{km.s^{-1}},
\end{align}
and we show it as a green square.
The  lower values  of $H_0\rd$  are consistent  with these  results,
while the higher value, inconsistent together, are also inconsistent
with Planck.

\begin{table}
  \centering
    \caption{\label{tab:H0rd}%
    $H_0\rd$ derived  model-independently from
    Eq.~\eqref{eq:H0rdB} and \eqref{eq:H0rdB}.
    The second column  is the median value of the  weighted average of
    the measurements at  $z=0.32$ and $0.57$, the third  column is the
    error given by the supernovae,  defined by the minimum and maximum
    values  of  $H_0\rd$ among  all  reconstructions,  and the  fourth
    column shows the errors given by the BAO.
  }
  \begin{tabular}{llll}
    \toprule
    $H_0\rd$ &
    Median &
    Error (SN) &
    Error (BAO) \\
    &
    (\si{km.s^{-1}}) &
    (\si{km.s^{-1}}) &
    (\si{km.s^{-1}})\\
    \midrule
    Method A &
    \num{10712.37} &
    $^{+837.90}_{-871.45}$ &
    \num {+-378.38}\\
    Method B &
    \num{10033.20} &
    $^{+333.10}_{-371.81}$ &
    \num{+-128.19}\\
    \bottomrule
  \end{tabular}
\end{table}

\begin{table}
  \centering
  \caption{\label{tab:rd}%
    Estimation of \rd\ assuming $H_0$ from R16 and R15.
    The errors are defined in the same way as table~\ref{tab:H0rd}. 
  }
  \begin{tabular}{llll}
    \toprule
    \rd &
    Median  &
    Error (SN) &
    Error (BAO+$H_0$) \\
    &
    (\si{Mpc}) &
    (\si{Mpc})&
    (\si{Mpc}) \\ 
    \midrule
    R16, Method B&
    \num{137.40} &
    $^{+4.56}_{-5.09}$&
    \num{+- 3.88}\\
    R15, Method B &
    \num{142.11} &
    $^{+4.72}_{-5.27}$&
    \num{+- 5.69} \\
    \bottomrule
  \end{tabular}
\end{table}

  We then  calculated the  weighted average of  $H_0\rd$ over  the two
  redshifts for each reconstruction for  methods A and B, and reported
  the median value in table~\ref{tab:H0rd}.
  The third column  is the error due  to the smoothing of  the SN data,
  given by the minimal and  maximal $H_0\rd$ over all reconstructions,
  and the fourth column shows the errors due to BAO.
  The   results   from   both   methods  are   consistent   with   the
  flat-\lcdm\ model (eq.~\ref{eq:Planck15}).
  Method  B  has smaller  error-bars  and  does not  assume  flatness,
  therefore we chose it as our main result.
  
  Using an external measurement of $H_0$, we can then estimate \rd.
  Table~\ref{tab:rd} shows our  estimation of \rd\ from  method B, for
  the two values of  $H_0$ from Riess \cite{2016arXiv160401424R} (R16)
  and   Rigault (R15) \cite{2015ApJ...802...20R}.
  The medians  and errors are  estimated in the  same way as  those of
  $H_0\rd$.
  The values of \rd\ obtained from $H_0$ from R15 are fully consistent 
  the \lcdm\ best-fit value of
  $\rd\  = \SI{147.41+-0.30}{Mpc}$  \citep{2016A&A...594A..13P}, while
  with $H_0$ from R16, the results are consistent within $2\sigma$.
  In  spite  of  the  large  error-bars, our  approach  gives  a  fully
  model-independent  estimation   of  $\rd$  where  we   even  used  a
  non-parametric   approach  in   reconstruction   of  the   expansion
  history. We should emphasize here that the tight constraints on $H_0
  \rd$  for the  \lcdm\  model are  due to  parametric  nature of  data
  fitting. Otherwise,  our reconstructions all have  better likelihood
  to the data than the best-fit 
  \lcdm\ model.

\subsection{Curvature test: $\Theta(z)$ and $\Ok(z)$} 
\label{sec:ktest}

The      $\Ok      (z)$     diagnostic      \cite{2008PhRvL.101a1301C,
  2010PhRvD..81h3537S,  2014PhRvD..90b3012S, 2016IJMPD..2530007B}  was
introduced to  test the FLRW  metric as well  as the curvature  of the
Universe.  
\begin{align}
  \Ok(z) &= \frac{(h(z)\DD'(z))^2-1}{\DD^2(z)}\label{eq:Ok}\\
  \intertext{
    In a FLRW metric, $\Ok(z)$ is constant and equal to \Omk.
    In a flat-FLRW metric, one has $\Ok(z)=0$, or}
  \label{eq:theta}
  \Theta(z) &\equiv   h(z)\DD'(z) = \frac{H(z)}{H_0}\DD'(z)= 1. 
\end{align}
Therefore, \Ok\  can be seen as  a curvature test, and  (if consistent
with a constant  value), gives the curvature  density parameter, while
$\Theta(z) = 1$ is a yes/no test to flat-FLRW.

By noticing  that $h(z) = H(z)\rd  / H_0\rd$, one can  combine the BAO
data with $H_0\rd$ calculated from eq.~\eqref{eq:H0rdB} in \S~\ref{sec:H0rd}. 
Both diagnostics can then be rewritten fully in terms of observables, as
\begin{align}
  \label{eq:Theta2}
  \Theta(z) &= \frac{1+z} c \left(H(z)\rd \frac{\dang(z)}\rd\right)
  \left(\frac{\DD'(z)}{\DD(z)}\right), \\
  &=  F_\mathrm{AP}(z) \left(\frac{\DD'(z)}{\DD(z)}\right)\\
  \Ok(z) & = \frac{\Theta^2(z)-1}
     {\DD^2(z)}, \label{eq:Ok2}
\end{align}
where   \DD\   and   $\DD'$   are   obtained   by   smoothing   method
(\S~\ref{sec:smoothing}),   $H(z)\rd$ and $\dang(z)/\rd$  are given
by    the     radial    and     tranverse    mode    of     the    BAO
\cite{2016MNRAS.457.1770C}, and $F_\text{AP}(z)  = (1+z) \dang(z)/ c(z)$
is the Alcock-Pazcynski anisotropy parameter.
We stress that in eq.~\eqref{eq:Theta2}, the first parenthesis depends
on the  BAO measurements, while the  second one depends on  the smooth
supernovae data.
In this new formulation, both  statistics  thus come  entirely  from
the BAO  and SN data,  unlike in \cite{2010PhRvD..81h3537S}  where the
authors also used $H_0$, and are fully model-independent.

\begin{figure}
  \centering
  \includegraphics[width=\textwidth]{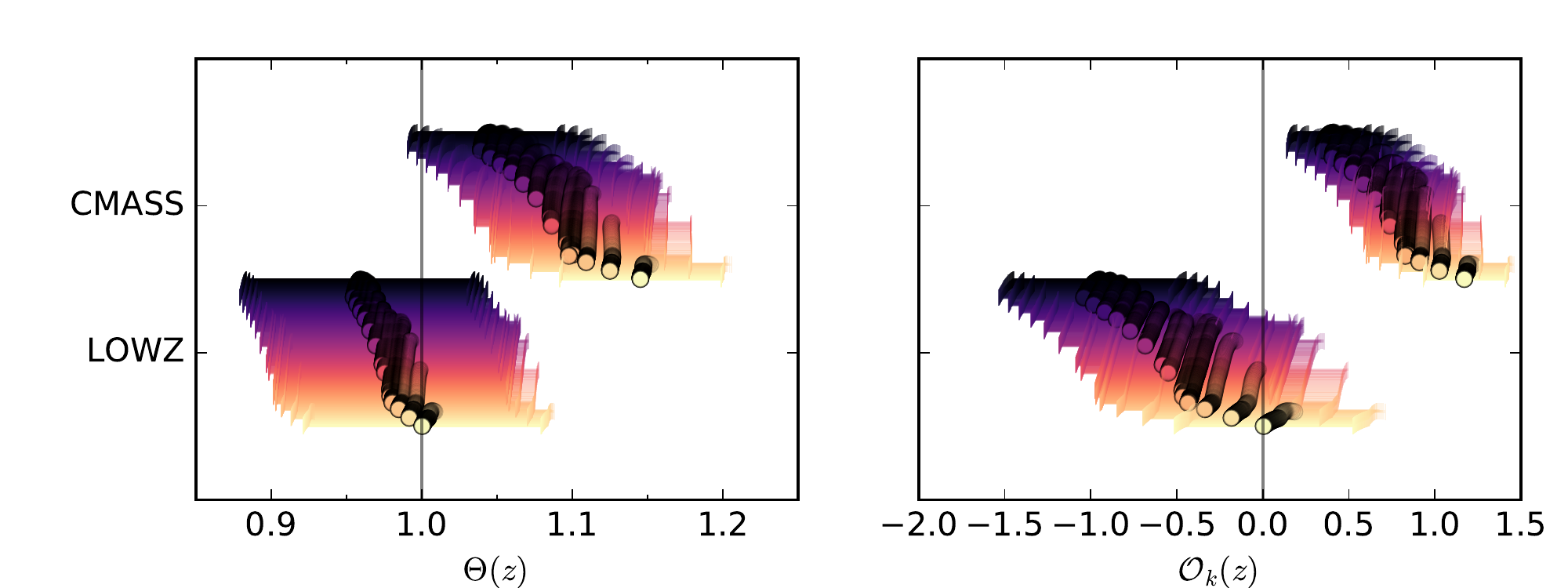}
  \caption{\label{fig:ktest} 
    Curvature    test.      Left:    $\Theta(z)    =     (1+z)/c    \,
    (H(z)\dang(z))(\DD'(z)/\DD(z))$;
    right: $\Ok(z)$ (eqs.~\ref{eq:Ok}).  
    Both statistics use $H(z)\rd$ and $\dang(z)/\rd$ from BAO
    measurements, and $\DD(z)$ and $\DD'(z)$ are obtained from the JLA
    supernovae data.  
  }
\end{figure}

The    left-hand   panel    of    Fig.~\ref{fig:ktest}   thus    shows
$\Theta$ at the LOWZ and CMASS redshifts (0.32 and 0.57).
Similarly  to Fig.~\ref{fig:H0zboss},  each  point  corresponds to  one
reconstruction      of      the       expansion      history      with
$\chsq<\chsq_\mathrm{ref}$,  while  the  error-bars  are  coming  from
$H(z)\rd$ and and $\dang(z)/\rd$,  taking into account the correlation
between $\dang$ and $H$.  
The right-hand panel  of Fig~\ref{fig:ktest} shows $\Ok(z)$
at the LOWZ and CMASS redshifts.
Here again, points in the same colour should be compared together.
At $z=0.32$, $\Theta$ is consistent with one for each reconstruction.
However, at $z=0.57$,  reconstructions with a lower  rate of expansion
history (yellow) show some tension with $\Theta=1$, while
reconstructions with higher rate of expansion history (dark) are
still consistent.
Regarding \Ok, the reconstructions with higher expansion history show
better consistency with a flat-FLRW universe at both redshifts. 
We should note that any inconsistency can be interpreted in two ways:
(1) the metric is indeed not FLRW, and therefore $\Ok$ is not equal to
\Omk.
(2) There  is some  systematics in  the supernovae  and/or in  the BAO
data.
If the metric is flat-FLRW, then $\Theta=1$ translate into
\begin{equation}
  H(z) \dang(z) = \frac {1+z} c \frac {\DD(z)} {\DD'(z)}.
\end{equation}
Therefore, an  inconsistency between  the two  datasets can  break the
equality.
One should  note that it  is possible  to test this  equality relation
using  only the  observables of  BAO and  supernovae data  taking into
account all correlations.

It is interesting to notice we have reached  an era where such litmus
tests can be directly  applied to the data to test  the pillars of the
concordance model with a reasonable precision.
Future  surveys will  be  able to  bring down  those  errors, thus  to
improve  the  constrain  on  the  flatness  without  parametric  model
assumptions.  

\section{Discussion and conclusion}
\label{sec:ccl}

  Using the most recent BAO and supernovae data (BOSS DR 12 and JLA), we
  estimated  $H_0\rd$, where  \rd\ is  the sound  horizon at  the drag
  epoch, in a model-independent way.
  We calculated
  $H_0 \rd = (\num{10033.20} 
  ^{+333.10}_{-371.81} \,(\text{SN}) \num{\pm128.19} \,(\text{BAO}))$ 
  \si{km.s^{-1}}, which is
  consistent with the value from Planck
  $H_0\rd = \SI{9944.0+-127.4}{km.s^{-1}}$ for the concordance model.  
  Without  any  assumption  from  the  CMB, and  only  using  BAO  and
  supernovae  data,  out  results  agree  with  the  (model-dependent)
  best-fit value $H_0\rd$ from Planck.

  Using    two   astrophysical    values   of    $H_0$,   we    derive
  \rd\ without using any parameterization and in a model-independent way.
  While  the  error-bars  are  large, the  results  using  the  Rigault
  measurement of $H_0$ are consistent  with the Planck estimated value
  of the concordance model, while those using
  the Riess measurements show some tension with Planck.

  We then tested the FLRW metric  and the flatness of the Universe at
  redshifts 0.32 and 0.57. 
  We used \Ok\ and introduced a new diagnostic $\Theta(z) = h(z)\DD'(z)$ 
  to quantify departure from flatness and FLRW.
  For a flat-FLRW Universe, $\Theta(z)=1$.
  Our  statistics  can  be  fully  written  in  terms  of  observables
  ($H(z)\rd$, $\dang(z)/\rd$, $\DD(z)$ and $\DD'(z)$), and our test is
  therefore fully model-independent.
  We  found   some  hints that there might be some inconsistency with  flat-FLRW
  (Fig.~\ref{fig:ktest}), which  may point to some systematics in the BAO or/and
  SN data or toward an actual departure from flatness.
  However, the results are still consistent with data fluctuations, therefore, 
  better data are needed to conclude. 
  Interestingly, the quality of the data has reached a level where we
  can consider such direct litmus tests very much plausible.  

  We  should  note here  that  there  have  been  many articles  in  the
  literature  using  $H(z)$  derived  from  age  of  passively  evolving
  galaxies     through     the      cosmic     chronometers     approach
  \cite{2003ApJ...593..622J, 2005PhRvD..71l3001S,   2009MNRAS.399.1663G,
    2010JCAP...02..008S,2016JCAP...05..014M}. However,  to derive $H(z)$
  through this  approach we  have to make  strong assumptions  on galaxy
  evolution characteristics.  Therefore in our  work we did not  use any
  $H(z)$ data  derived from age  of passively evolving galaxies  and our
  analysis is solely based on the BAO data from the BOSS survey.

Our internal  consistency tests show some  interesting results. While
we  expect to  see  a  proper consistency  between  derived values  of
$H_0\rd$ 
using  LOWZ  and  CMASS  data considering  both  $d_A(z)$  and  $H(z)$
observations,  we  realized  some considerable  tensions.  Having  the
expansion  history directly  derived from  supernovae data  we noticed
that   only   by   considering   lower   values   of   $H_0\rd$   (around
\SI{e4}{km.s^{-1}} and also considering larger expansion 
history rate (darker lines in  Fig.~\ref{fig:Dall} for $D(z)$), we can
have consistent results between all measurements. This is particularly
important looking at the results from CMASS data. 
This  somehow   supports  the  cosmological  parameters   from  Planck
concordance model  cosmology (lower  $H_0$ and higher  $\Omm$) without
using the cosmic microwave background data. 
Future surveys, such as DESI, will measure $H(z)\rd$ at several redshifts with
smaller uncertainties.  Using these values,  we can test  the flatness
and the  metrics in a wider  range and with much  higher precision and
accuracy. 

\section*{Acknowledgement}

The  authors thank  Eric Linder  for various  useful discussions,  and
Changbom Park for his comments.
A.S.  would like to acknowledge the support of the
National Research Foundation of  Korea (NRF-2016922914). This work was
supported   by   the   National  Institute   of   Supercomputing   and
Network/Korea  Institute of  Science and  Technology Information  with
supercomputing resources including technical support.

\appendix

\end{document}